\documentstyle[12pt,epsfig]{article}
\date{ \today }
\begin{document}
\pagestyle{plain}
\topmargin -1.3cm
\textheight 21.6cm
\textwidth 15.0cm
\def\baselinestretch{1.2}
\tolerance=3000
\baselineskip=20pt

\begin{center}
{\large \bf Mass and charge identification of fragments detected with
the Chimera Silicon-CsI(Tl) telescopes}\\
N.~Le~Neindre$^{(a)}$, M.~Alderighi$^{(b)}$, A.~Anzalone$^{(c)}$,
R.~Barn\`a$^{(d)}$, M.~Bartolucci$^{(e)}$, I.~Berceanu$^{(f)}$,
B.~Borderie$^{(g)}$, R.~Bougault$^{(h)}$, M.~Bruno$^{(a)}$,
G.~Cardella$^{(i)}$, S.~Cavallaro$^{(c)}$, M.~D'Agostino$^{(a)}$,
R.~Dayras$^{(j)}$, E~.De~Filippo$^{(i)}$, D.~De~Pasquale$^{(d)}$,
E.~Geraci$^{(c)}$, F.~Giustolisi$^{(c)}$, A.~Grzeszczuk$^{(l)}$,
P.~Guazzoni$^{(e)}$, D.~Guinet$^{(m)}$, M.~Iacono-Manno$^{(c)}$,
A.~Italiano$^{(d)}$, S.~Kowalski$^{(l)}$, A.~Lanchais$^{(a)}$,
G.~Lanzano'$^{(i)}$, G.~Lanzalone$^{(i,g)}$, S.~Li$^{(n)}$, S.~Lo
Nigro$^{(i)}$, C.~Maiolino$^{(c)}$,  G.~Manfredi$^{(e)}$,
D.~Moisa$^{(f)}$, A.~Pagano$^{(i)}$, M.~Papa$^{(i)}$,
T.~Paduszynski$^{(l)}$, M.~Petrovici$^{(f)}$,
E.~Piasecki$^{(o)}$, S.~Pirrone$^{(i)}$, G.~Politi$^{(i)}$,
A.~Pop$^{(f)}$, F.~Porto$^{(c)}$, M.~F.~Rivet$^{(g)}$,
E.~Rosato$^{(k)}$, S.~Russo$^{(e)}$, S.~Sambataro$^{(i,+)}$,
G.~Sechi$^{(b)}$, V.~Simion$^{(f)}$, M.~L.~Sperduto$^{(c)}$,
J.~C.~Steckmeyer$^{(h)}$, C.~Sutera$^{(i)}$, A.~Trifir\`o$^{(d)}$,
L.~Tassan-Got$^{(g)}$, M.~Trimarchi$^{(d)}$, G.~Vannini$^{(a)}$,
M.~Vigilante$^{(k)}$, J.~Wilczynski$^{(p)}$, H.~Wu$^{(n)}$,
Z.~Xiao$^{(n)}$, L.~Zetta$^{(e)}$, W.~Zipper$^{(l)}$ \vspace{.5cm}
\end{center}
\noindent
 {\small \it
$^{(a)}$ INFN and Dipartimento di Fisica Universit\`a di Bologna, Italy,\\
$^{(b)}$ Istituto Fisica Cosmica, CNR and INFN Milano, Italy,\\
$^{(c)}$ Laboratorio Nazionale del Sud INFN and Dipartimento di
Fisica Universit\`a di Catania,\\
$^{(d)}$ INFN and Dipartimento di Fisica Universit\`a di Messina,\\
$^{(e)}$ INFN and Dipartimento di Fisica Universit\`a degli Studi, Milano,
Italy,\\
$^{(f)}$ Institute for Physics and Nuclear Engineering, Bucharest, Romania,\\
$^{(g)}$ IPN, IN2P3-CNRS and Universit\'e Paris-Sud, Orsay, France\\
$^{(h)}$ LPC,ISMRA and Universit\'e de Caen, France \\
$^{(i)}$ INFN and Dipartimento di Fisica Universit\`a di Catania,\\
$^{(j)}$ DAPNIA/SPhN - CEA Saclay, Gif sur Yvette Cedex, France\\
$^{(k)}$ INFN and Dipartimento di Fisica Universit\`a di Napoli, Napoli,
Italy,\\
$^{(l)}$ Institute of Physics, University of Silesia, Katowice, Poland,\\
$^{(m)}$ IPN, IN2P3-CNRS and Universit\'e Claude Bernard, Lyon, France,\\
$^{(n)}$ Institute of Modern Physics Lanzhou, China,\\
$^{(o)}$ Institute of Experimental Physics, University of Warsaw, Poland,\\
$^{(p)}$ Institute for Nuclear Studies, Otwock-Swierk, Poland\\
$^{(+)}$ Deceased
 }

\vspace{.5cm}

\small
\begin{abstract}
Mass and charge identification of charged products detected with
Silicon-CsI(Tl) telescopes of the Chimera apparatus is presented.
An identification function, based on the Bethe-Bloch formula, is
used to fit empirical correlation between $\Delta E$ and $E$  
ADC readings, in order to determine, event by event, the atomic 
and mass numbers of the detected charged reaction products 
prior to energy calibration.\vspace{.5cm}

\end{abstract}

\normalsize

\section{Introduction}
In the last years several new detectors for charged particle
identification, with large solid angle coverage and high
geometrical efficiency, have been built in order to investigate
heavy ions reactions at intermediate energies ($10$A MeV-$\sim
1$A GeV)~\cite{apparati,chimera}.

These experimental devices give possibilities for simultaneous
measurement of quantities related to energy, emission angle,
atomic number and mass number of nearly all the charged reaction
products. A very rich information can be extracted from these
experimental studies, in particular if the mass resolution of the
apparatus is high. Indeed, isotope yields have proven to be useful
observable for studying the heavy-ion reaction mechanisms at
intermediate and high energies~\cite{trautmann}. Production
yields of isotopically resolved nuclear particles and fragments
can provide answers to the question of mutual stopping and
subsequent equilibration of the collision partners. On the
condition that equilibrium is reached, they allow to extract the
corresponding thermodynamical variables, such as the temperature
of the hot pieces of nuclear matter formed in the collision,
their density and possibly the entropy.

The necessary step before the data analysis is the calibration of
the measured signals. However, due to the fact that different
detectors (ionization chambers, microstrips, semiconductors,
scintillators) can be used, due to the rich variety of nuclear
species produced in the reaction in a wide energy range and to
the large number of telescopes covering the laboratory solid
angle, this preliminary step is quite man-power and time
consuming.

Another difficulty is that not necessarily all the detectors of 
a multi-modular $\Delta E-E$ telescope have response linear in energy. 
The last-generation $4 \pi$ devices~\cite{apparati,chimera}
are multi modular $\Delta E-E$ telescope systems, in which the 
residual-energy detectors are usually made of the CsI(Tl) scintillator. 
This choice is dictated by its relatively high stopping power, no limitation 
in geometrical shapes, negligible radiation damage, low cost, and 
good resolution. Since the light output of a scintillator depends 
both on the energy deposited in the cristal and on the atomic and 
mass numbers of the incident ion, the identification in mass and charge 
has to be done prior to energy calibration~\cite{cesimcs}.

For the charge identification, semi-automatic~\cite{paolocalib} and 
automatic~\cite{guazzoni} techniques have been set in recent years, 
but no recent literature~\cite{old} is available for the mass
identification, usually performed through graphical
cuts~\cite{prctrieste} around each A-line in the $\Delta E-E$
scatter plot. This method, however, has the disadvantage that the
number of contour lines that can be drawn is limited by the
statistics and extrapolation to rare isotopes cannot be
performed. Part of the physical information for isotopes
populated with low statistics is then lost. In addition,
measurements with neutron rich/poor beams and targets could lead
to not negligible shifts of the mass distributions with respect
to stable isotopes, making difficult the A-labeling of the
graphical cuts.

A fast and reliable method to assign the mass and charge of the
detected ions is therefore highly desirable.

In this paper we present a very effective mass and charge
identification procedure which, as compared with other methods,
considerably saves time without loss of precision. The accuracy
of the proposed method is checked by comparing the mass
distributions obtained with this procedure with original mass
distributions in model-generated samples.

We present here the application of our identification procedure to
data collected in the first experiments performed by the Reverse
collaboration with the forward part of the Chimera
apparatus~\cite{cris2000}. Mass and charge distributions obtained
through more usual methods, like graphical cuts or particle
identification functions, are compared with those obtained from
our procedure.

\section{ Mass and charge particle identification function}
We recall here the main principles leading to the used
identification function. A more detailed description of the
successive modifications of the Bethe-Bloch formula, to take care
of the experimental distortions, can be found in~\cite{tg}.

Specific energy loss (${{dE}\over {dx}}$) of a charged particle
in matter depends , as described by the classical Bethe-Bloch
formula~\cite{bethe}, on the mass $A$, charge $Z$ and energy $E$
of the incident ion and of the density and atomic number of the
absorbing medium.  For light ions with incident energy high
enough to approximate the effective charge by the charge $Z$ of
the ion, the specific energy loss is:
\begin{equation}
\frac{dE}{dX} = \frac{Z^2}{f(E/A)} \label{eq1}
\end{equation}
Analytical reductions~\cite{tg} of this expression lead to:
\begin{equation}
\Delta E = \left[ E^{\mu +1} + (\mu +1) Z^2 A^{\mu} \Delta X
\right]^ {\frac{1}{\mu +1}} - E \label{eq5i}
\end{equation}
in the case of particles detected in a $\Delta E-E$ telescope.
$\Delta X$ is the thickness of the first detector, where the ion
deposits an energy $\Delta$E. In the second detector the ion is
stopped and releases an energy $E$. To obtain eq.(\ref{eq5i}) from
eq.(\ref{eq1}) the hypothesis that $f(E/A)$ is a power-law
\begin{equation}
f(E/A) = (E/A)^{\mu} \label{eq4}
\end{equation}
with exponent $\mu \approx 1$ has been made~\cite{goulding}.

Eq.(\ref{eq5i}) is the basic formula to build particle
identification functions $(p.i.f.)$ for charge
identification. For instance, if we aim to identify the charge Z
of a detected particle/fragment from the measured $\Delta E, E$
signals, we can calculate a not calibrated measure of $Z$
($p.i.f.$), which includes some unknown constants (like for
instance the thickness of the $\Delta E$ detector and the exact
relationship between the mass and the charge):
\begin{equation}
 p.i.f. = \left[ \left( \Delta E + E \right)^{\mu +1} - E^{\mu +1}
\right]^\frac{1}{(\mu +2)} \label{pif}
\end{equation}
with the assumption $A = 2 \ Z$.

However, it is experimentally well known that it is quite
difficult, by managing  the only parameter $\mu$, to find  a
unique $p.i.f.$  able to linearize the $\Delta E-E$ correlation of
each used telescope, in the whole range of residual energies and
for a wide range of charges, as usually observed in heavy ion
reactions.

Modifications to eq.(\ref{eq5i}) are therefore needed, since data
deviate from the expected behaviour for several reasons:
\begin{itemize}
\item when the residual energy becomes low, the atomic charge is no
longer equal to Z, especially for heavier elements;
\item in experiments where the ion is stopped in a
scintillator, the residual energy signal is not linear with the
released energy. For a scintillator indeed, the light output
response depends also on A and Z~\cite{cesimcs};
\item when $\Delta E$ is measured with a Silicon
detector, the pulse height defect influences the Silicon detector
response for high Z-values.
\end{itemize}

To evaluate mass and charge, before energy calibration, an
extension of eq.(\ref{eq5i}), taking into account all the
aforementioned problems has been proposed in Ref.~\cite{tg}. This
formula performs a decoupling of the $\Delta E-E$ correlation at
low, intermediate and high energies, by introducing some free
parameters and a phenomenological term, which takes care for the
transition from low to high energies.
\begin{equation}
\Delta E = \left[ (g E)^{\mu + \nu + 1} + \left( \lambda Z^\alpha
 A^\beta \right)^{\mu + \nu + 1} + \xi Z^2 A^\mu (g E)^\nu
 \right]^ {\frac{1}{\mu + \nu + 1}} - g E \label{eq8}
\end{equation}
where $\lambda, \mu, \alpha, \beta, \nu$ and $\xi$ are free
parameters, related to the characteristics and non-linear effects
of the $\Delta E$ and $E$ detectors. $g$ accounts for the ratio
of the electronic gains of the $\Delta E$ and $E$ signals (QDC or
ADC channels). Eq.(\ref{eq8}) contains 7 free parameters in case
the subtraction of $\Delta E$ and $E$ pedestals has already been
performed.

Eq.(\ref{eq8}) reduces to eq.(\ref{eq5i}) if $g=1$, $\nu = 0$,
$\xi = 0$, $\alpha = 2/(\mu +1)$, $\beta= \mu /(\mu +1)$. In this
case the parameter $\lambda$ results equal to $[(\mu + 1) \Delta
X]^\frac{1}{\mu+1}$. Eq.(\ref{eq8}) (see Ref.~\cite{tg}) provides
the same behaviour as Eq.(\ref{eq5i}) at vanishing and at high
residual energies.

In Ref.~\cite{tg} the application of eq.(\ref{eq8}) has been shown
to be powerful for charge identification, in a wide range of
products charge and for different $\Delta E-E$ types of
telescopes. This has also been found to be valid in the case of
new designed gas detectors as microstrips~\cite{garfield}.

As far as the mass identification is concerned, eq.(\ref{eq8})
has been applied in Ref.~\cite{tg} to a limited number of
isotopes, since the data there presented were collected with
devices providing mass resolution only for light fragments ($Z\le
4$).

In the following we will show the application of eq.(\ref{eq8}) to
the Chimera~\cite{chimera} Si-CsI(Tl) telescopes, where the mass
resolution is high up to $Z \le 10$.

\section{Identification procedure}
The used identification procedure, consists of two steps:
\begin{itemize}
\item
We sample on the $\Delta E-E$ scatter plot several points on the
lines of well defined isotopes ($He, Li, Be$ and $B$). In
experiments, some isotopes can be easily recognized, due to their
abundance ($^{4}He, ^{7}Li, ^{11}B$) or separation from other
masses ($^{7}Be, ^{9}Be$).

The charge, mass, $\Delta E$ and $E$ signals of the sampled
points are put in a table. A minimization routine~\cite{minuit}
determines the parameters $g$, $\lambda, \alpha, \beta, \mu$,
$\nu$ and $\xi$, giving the best agreement between the whole
sample and the correlation provided for each A and Z by
eq.(\ref{eq8}). The sum of the squared distances between the
sampled and calculated values is minimized.

This procedure is performed for each used telescope (688 for the
case of the Reverse experiments) and a map is built containing
the identification number of the telescope and its characteristic
parameters $\lambda, \mu, \alpha, \beta, \nu, \xi$  and $g$.

\item We perform the event by event identification.

In each event, each detected particle/ion is identified in mass
and charge by a two-step process, by minimizing the distance of
the measured $\Delta E$ and $E$ signals with respect to the values
provided by eq.(\ref{eq8}) with the parameters $g$, $\lambda,
\alpha, \beta, \mu$, $\nu$ and $\xi$ read from the map built in
the previous step. The 2-dimensional vector ($\Delta E$, $E$) is
then replaced by the 4-dimensional vector ($\Delta E$, $E, Z, A$)
for subsequent analyses.

To identify mass and charge of the detected charged products a
two-step process is needed, since eq.(\ref{eq8}) is not
analytically solvable. The first step is to find the charge $Z$
(simply assuming $A = 2 Z$), by looking for the value of $Z$
giving the shortest distance between the experimental $\Delta E$
and the energy loss given by eq.(\ref{eq8}) at the residual
energy $E$. After the charge $Z$ has been identified, the
procedure is repeated, by solving eq.(\ref{eq8}) with respect to
$A$.
\end{itemize}

This procedure is an improvement of the method developed in
Ref.~\cite{paolocalib}, extensively used in the Multics
experiments for charge identification (see for instance
Ref.s~\cite{prctrieste,comparisonmcs,paper_qp}) when cocktail
beams were not available.

The improvement consists not only in the simultaneous event by
event mass and charge identification, but mainly in the use of the
analytical form of the $\Delta E-E$ correlation eq.(\ref{eq8})
together with the map of parameters characterizing the individual
response of each telescope. Deviations of the Silicon detector
thickness with respect to the values provided by the factory as
well as intrinsic scintillation efficiency of the CsI(Tl)
crystals are taken into account by the map parameters.
Consequently, the interpolation or extrapolation to values of $A$
and $Z$ not sampled or emitted with low probability should be
under control.

This will be verified in the following Sections.

\section{Check of the procedure through model events}
As an example of the identification procedure, we first show the
application of our method to a set of ideal events, where not
only the signals $\Delta E$ and $E$ are known, but also the mass
number $A$ and the charge number $Z$. The comparison between the
original values of ($Z$, $A$) and the reconstructed ones allows
to evaluate the capability and the limits of the procedure.

The {\it ideal} events used to check the identification function
(\ref{eq8}) are generated by the statistical multifragmentation
model SMM~\cite{bondorf}. This model reproduces in great detail
static and dynamic experimental observables, like charge, mass
and kinetic energy distributions~\cite{prctrieste,comparisonmcs,
comparisonindra}. We analyze here events of the decay of a $Au$
source, with excitation energy in the range $1-8 \ A MeV$ and
density one third of the normal density~\cite{paper_qp}.

For each event, from the mass number, charge number and kinetic
energy of the charged decay products we compute their $\Delta E$
(energy loss in a 280 $\mu m$ Silicon detector) and the residual
energy $E$.

In Fig.~1 we show the $\Delta E-E$ plots for $Z=2$, $Z=3$ and $Z
\ge 4$ (panels a, b, and c), respectively). The stars in Fig.~1
represent the sampled points, used to calculate the parameters of
eq.(\ref{eq8}). From the 7 parameters fit of the sample (165
points, step 16 MeV in the residual energy), we obtain a
description of the $\Delta E-E$ correlation under study,
characterized by an average distance between the sampled points
and the values calculated through eq.(\ref{eq8}) of $\approx 0.1
\ MeV$.

Since we are dealing with energies, the ratio of the gains $g$
results $\approx 1$. The parameter $\nu$ results nearly zero and
$\alpha$ and $\beta$ satisfy the relationships $\alpha = 2/(\mu +
1)$, and $\beta = \mu/(\mu + 1)$, as expected for an ideal
detector.

We compare now the original mass distribution of the model, for
isotopes not used for the fit, with the one calibrated through our
identification procedure. We want indeed to estimate how far we
can extrapolate above the highest charge contained in the sample 
($Z = 6$), still having a reliable reproduction of the A-distributions.

In Fig.~2 we compare the $A$-distributions for charges not
included in the fit: $Z=1$ and $Z=8-12$. Histograms are the
original yields of the model, symbols are values
obtained from eq.(\ref{eq8}). For $Z = 1$, $Z = 8$ and $Z = 9$ 
the calculated yields are in perfect agreement with the model. 
For $Z \geq 10$ discrepancies appear between original and
reconstructed mass yields. For instance for $Z=10, A = 20$ 
and for $Z = 12, A=26$ the deviation is about 15\%.

As far as the charge identification is concerned, over the whole
set of considered charged products ($1\leq Z \leq 12$),
eq.(\ref{eq8}) has been found to fail in the charge determination
(by one charge unit) with a probability smaller than $3 \
10^{-5}$.
We have also found that, when interpolating between sampled
$\Delta E-E$ lines, the calculated mass yields are in perfect
agreement with the model ones.

The results shown in Figures 1 and 2 have been obtained for an
ideal telescope, with perfect resolution. Now we consider the
modifications induced by finite energy resolution. We perform this
check in the case of an energy resolution 2\% for the $\Delta E$
detector and 4\% for the stopping detector (much worse than
those  observed for the Chimera detectors). The chosen values of
the resolutions still allow to resolve the isotope lines.

By repeating the analysis shown in Figure 2, we have found the
same results as before for charge determination and mass
interpolation. As far as mass identification for charges above the
highest charge contained in the sample, the agreement between true and 
calculated mass yields slightly worsens with respect to the case 
of a "perfect" detector. Indeed for $Z=9$ ($A = 20$) calculated values 
differ by 4\% with respect to the original ones, for $Z=10, A = 20$ 
and for $Z = 12, A=26$ the deviation is about 25\%.

These results establish the confidence level for mass
identification of charges not included in the sampled set.
We conclude that the proposed procedure is suitable to reliably extrapolate 
masses up to  3 charges above the highest charge contained in the sample, 
even when considering $\Delta E-E$ correlations spread by the detector 
resolution.

\section{Experimental results}
In this Section we show the results of the identification
procedure applied to experimental data. The data have been
collected, for the reactions $^{124}$Sn+$^{64}$Ni,
$^{112}$Sn+$^{58}$Ni and $^{124}$Sn+$^{27}$Al at 35 A MeV
incident energy, in experiments performed at the Superconducting
Cyclotron of LNS (Catania) by the Reverse
collaboration~\cite{bologna2000}.  The forward part
($\theta_{lab} \le 30^o$) of Chimera array was used for the
experiments. In this configuration, 688 telescopes made of
$\Delta E$ Silicon detectors 200-300 $\mu m$ thick (depending on
$\theta_{lab}$) and CsI (Tl) stopping detectors were used.

The energy signal of the silicon detector was obtained by a
standard  spectroscopic line, made by a fast low noise charge
sensitive preamplifier (PAC), followed by a main spectroscopic
amplifier ($\sim 0.75 \ \mu s$ shaping time).

The light output of the crystal was collected by a $20 \times 20
mm^2$ photodiode coupled with a low noise $45 \ mV/MeV$ PAC.
The PAC output signal was shaped by a spectroscopic amplifier ($2 \mu s$
shaping time) and the shaped signal was stretched in order to
avoid any time jitter in the  digital conversion performed by a 64
channels VME single gate QDC.

The energy resolutions of the silicon detectors and the CsI(Tl)
crystals were quoted by measuring the elastic scattering of
different ion beams, delivered by the Tandem and the Cyclotron
accelerators of the LNS in Catania, impinging on a thin ($\sim
100 \mu g /cm^2$) Au target. The typical energy resolution of a
Chimera telescope resulted $< 1\%$ for the Silicon detector and
about 2\% for the CsI(Tl). As an example, with beams of $^{58}$Ni
at 15.5 A MeV the energy resolution (FWHM) was 0.5\% for the
Silicon detector and 1.5\% for the crystal, using the lowest
value of the gain (g=1 Volt/Volt) of the main amplifier. The
energy resolution of silicon detectors was also measured by
collecting $\alpha$-particles of a standard three peaks
radioactive source. Typical energy resolutions were $\sim 70 \
keV$ with g= 8 Volt/Volt and $\sim 200 \ keV$ with g= 2
Volt/Volt. In order to reduce possible distortions of the
electric field, the polarization bias of the Silicon detectors
was increased by 30\% with respect to the nominal one. As a
global result, in 95\% of the $\Delta E-E$ matrices a good
identification of the atomic number (up to a charge $Z = 50$ for
the most forward angle),  was obtained in the full dynamical
range of the experiment (see Fig.6). Finally, in the high gain
conversion range of  the QDC~\cite{chimera}, corresponding to a
dynamical range of about 120 MeV with a PAC sensitivity of 4.5
mV/MeV and g=2 Volt/Volt, a good identification of isotopes from
charge Z=3 up charge Z=8 was clearly achieved (see Fig. 3).

\subsection{Mass identification}
We start the analysis, by sampling a set of ($\Delta E, E$)
points (see Fig.~3). The isotopes chosen to
build the sample set are $^{4}He, ^{7}Li, ^{7}Be, ^{11}B, ^{13}C$.

As a second step we fit the data (122 points) with
eq.(\ref{eq8}), with 9 free parameters (both $\Delta E$ and $E$
are QDC channels, the pedestals have not been subtracted).

The resulting total 
$\chi^2 = \frac{1}{d.o.f.} \sum 
\frac{(\Delta E - \Delta E_{calc})^2}{errors^2}$ 
is $2.3$ (d.o.f. stands here for the degrees of freedom, 
number of the sampled points minus
the number of free parameters). The errors on the experimental
sampled points (channels) have been estimated to be about 10
channels, by sampling several times the same $\Delta E-E$
experimental matrix. All the sampled isotopes give comparable
partial contributions to the total $\chi^2$.

Finally, the event by event identification was performed. The
resulting isotopic distribution for individual charges from $Z=3$
to $Z=8$ is shown in Fig.~4.  In the same figure is also reported
the mass spectrum, obtained for charges from $3$ to $10$ (from
$^{6}Li$ to $^{23}Ne$ respectively). The quality
of the mass identification obtained from our procedure is remarkable. 
Indeed,
the {\it peak over noise} ratio of the mass distribution, even
when integrated over eight charges, remains similar to the one
obtained by looking at separated charges.

We have checked the accuracy of our procedure, by comparing the
yields obtained from eq.(\ref{eq8}) with those obtained with
graphical cuts performed on the A-lines, in the regions where
these lines are clearly distinguishable. From Table I it is
evident that all the mass yields are in agreement within the
statistical uncertainties.  In addition the yields obtained from
eq.(\ref{eq8}) resulted in very good agreement with gaussian
integrals of the isotope distribution of Fig.~4.

We want here to remark another advantage of our identification
procedure with respect to graphical cuts.

The {\it a priori} A-labeling  of the observed $\Delta E-E$
correlations (see Fig.~3) in the region of heavy ($Z>4)$ isotopes
could be disputable, in absence of reference beams sent on the
detectors and/or before energy calibration and comparisons to
energy-loss calculation. Indeed, noticeable shifts of the mass
distributions with respect to stable isotopes are
expected~\cite{baran} in experiments running with neutron
rich/poor beams and targets (as in the case of Reverse
experiments).

In the case of the adopted procedure quantitative {\it
reliability} tests can be performed by checking the partial
contribution of the sampled isotopes to the total $\chi^2$. For
instance, if the fit of the sampled points (shown in Fig.~3) is
performed, by assuming the mass of the sampled $Z=6$ isotope as
$A=12$ (instead of $13$), the corresponding partial contribution
rises to a value about $25$ (instead of $\sim 2$).

As an example of the rich variety of isotopes produced in
reactions with neutron rich/poor projectile and targets, we
compare in Fig.~5 the mass distributions measured for the
reactions $^{124}$Sn+$^{64}$Ni and $^{112}$Sn+$^{58}$Ni at 35
A MeV (respectively N/Z=1.41, 1.18).  The isospin of the entrance
channel reflects on large shifts of the mass distributions.
Neutron rich systems enhance the production of neutron rich
isotopes, as expected from theoretical calculations~\cite{baran},
mainly in the even charge isotopes. Study of heavy ion reaction
mechanisms are therefore feasible, to provide information on
thermodynamical variables of equilibrated hot sources formed in
the collision, such as their temperature and density.

\subsection{Charge identification}
As another example of the identification method we show in the
following the application of our procedure to identify the charge
of fragments collected by a forward detector ($\theta_{lab}=1^o$)
of the Chimera array.

Also in this case we sample a set of ($\Delta E, E$) points. We
build a set of sampled points for $2\leq Z \leq 9$, $Z = 15, 20,
30, 40$ and $50$, and we fit this sample with
eq.(\ref{eq8}). We obtain a set of parameters, associated to the
examined detector, reproducing the sampled points with a 
$\chi^2 = \frac{1}{d.o.f.} \sum 
\frac{(\Delta E - \Delta E_{calc})^2}{errors^2}$ 
equal to 0.45. All the observed $\Delta E-E$ Z-lines are well 
reproduced up to the whole Z=50 line (see Fig.~6).

In Fig.~7 we compare the corresponding charge distribution with
the one obtained by a particle identification function
(eq.(\ref{pif})) (upper insert panel of Fig.~6). The overall
agreement is very good. Some disagreement between the two
distributions is present for low values of atomic numbers, which
suffer of poor statistics and at very high values of the charge
number, for which the determination of the charge through the
p.i.f results less precise than at intermediate Z-values. Indeed
the {\it peak over noise} ratio ranges from a value larger than 5
(up to charges $\simeq 30$) to $\sim 4$ at Z=50. As a
consequence, the contamination among adjacent charges goes from
2\% to about 4\%.

\section{Conclusions}
We have presented a semi-automatic $Z$ and $A$ identification
procedure, tested through the analysis of model events with known
distributions. A very good agreement was obtained.

The experimental results have been verified by comparing the
obtained $A$ and $Z$ distributions with those resulting from
standard methods (graphical cuts or particle identification
functions). The results of these comparison resulted quite
satisfactory, and the uncertainties on the charge and mass
identification are clearly improved.

The advantages of the procedure may therefore be summarized as
follows:
\begin{itemize}
\item Beams of known mass, charge and energy
are not needed for the charge and mass identification of charged
products, detected with  $\Delta E-E$ telescopes with good energy
resolution.
\item If mass is not resolved, Z-identification is still possible
by substituting in eq.(\ref{eq8}) $A$ with $2 Z$ or with other
parameterizations commonly used in experiments~\cite{moretto,epax}.
\item The accuracy of the mass identification is very high, at least
comparable in precision with the graphical cuts in $\Delta E-E$
matrices, but clearly more powerful because our procedure saves a 
lot of time, especially in experiments involving many telescopes.
\item The use of a map of identification parameters, taking into account
the response of each telescope, makes possible the
extrapolation to A-regions where graphical cuts are not easy to
make, due to low statistics.
\item Possible drifts of the electronic circuits
can be diagnosed by controlling the constancy of the parameters
of the map during the sequence of runs throughout the whole
experiment.
\item The method ensures relatively fast, standardized
and reliable mass and charge identification for multi-telescope
systems.
\end{itemize}

\section{Acknowledgements}
The authors wish to thank  
R.~Bassini, C.~Boiano, C.~Cali', V.~Campagna, R.~Cavaletti, O.~Conti, 
M.~D'Andrea, F.~Fichera, N.~Giudice, A.~Grimaldi, N.~Guardone, S.~Hong,
P.~Litrico, G.~Marchetta, S.~Marino, D.~Nicotra, G.~Peirong, C.~Rapicavoli, 
G.~Rizza, G.~Sacca', S.~Salomone, S.~Urso, 
for the technical support during the experiment. The authors are also
grateful to L.~Calabretta and D.~Rifuggiato for the assistance in
delivering beams of very good timing quality.

This work was supported in part by the Italian Ministry of University and
Scientific Research under grants Cofin99 and by NATO grants
PST/CLG 976861.

\newpage
\begin{table}[h!]
\begin{center}
\caption{Comparison of the isotope yields obtained from graphical
cuts and from eq.(\ref{eq8})}
\begin{tabular}{| c | c | c | c |}
\hline
Isotope & Graphical Cuts & Eq.(\ref{eq8}) \\
\hline
$^{6}$Li &  864 & 865  \\
$^{7}$Li &  1950 & 1948   \\
$^{8}$Li & 547 & 520 \\
$^{7}$Be & 264 & 244  \\
$^{9}$Be & 1114 & 1080  \\
$^{10}$Be & 1173 & 1109  \\
$^{10}$B & 350 & 354    \\
$^{11}$B & 1673 & 1600  \\
$^{12}$B & 494 & 519  \\
$^{12}$C & 745 &  707 \\
$^{13}$C & 860 &  858   \\
$^{14}$C & 616 & 579  \\
\hline
\end{tabular}
\end{center}
\end{table}

\newpage
\begin{figure}[ht]
\protect\caption{\it Model
events: $\Delta E-E$ plots (colors from yellow to black follow a
logarithmic scale) for ions passing through a $280 \mu m$ Silicon
detectors. Stars at $Z=2, A=4$ (panel a)), $Z=3, A=7$ (panel b)),
$Z=4, A=7$, $Z=5, A=11$ and $Z=6, A=12$ (panel c)) are the
sampled points used for the fitting procedure. }
\end{figure}
\begin{figure}[ht]
\protect\caption{\it Model
events: Isotopic distributions obtained from the extrapolation of
eq.(\ref{eq8}) to isotopes not used in the fit. Lines represent
the model A-distributions, full symbols the values calculated
from eq.(\ref{eq8}). }
\end{figure}
\begin{figure}[ht]
\protect\caption{\it
Experimental $\Delta E-E$ scatter plot for a Chimera telescope
at  $\theta_{lab} =21^o$. Red symbols are the sampled points used
in the fitting procedure: $^{4}He$ (open squares), $^{7}Li$ (open
circles), $^{7}Be$ (full squares), $^{11}B$ (stars), $^{13}C$
(full points). }
\end{figure}
\begin{figure}[ht]
\protect\caption{\it
Experimental isotopic distributions obtained for charges $3-8$.
The bottom histogram represents the mass distribution for
charges $3-10$. }
\end{figure}
\begin{figure}[ht]
\protect\caption{\it
Experimental isotopic distributions obtained for charges $3-8$.
The grey histogram refers to the reaction $^{124}$Sn+$^{64}$Ni
and the hatched one to the reaction $^{112}$Sn+$^{58}$Ni. The
histograms have been normalized to the total area for each
charge.}
\end{figure}
\begin{figure}[ht]
\protect\caption{\it
Experimental $\Delta E-E$ scatter plot for a Chimera telescope at
$\theta_{lab} =1^o$. Lines show the values calculated from
eq.(\ref{eq8}). The insert panel shows the charge identification
obtained with the p.i.f. method (see text). }
\end{figure}
\begin{figure}[ht]
\begin{center}
\end{center}
\protect\caption{\it Experimental charge distribution for the
$\Delta E-E$ matrix of Fig.5.  Symbols show the values calculated
from eq.(\ref{eq8}). Lines show the yields calculated through a
particle identification function (see text). }
\end{figure}
\end{document}